\documentclass{elsart1p}
\usepackage{graphicx}
\usepackage{amssymb}
\usepackage{amsfonts}
\begin{document}
\begin{frontmatter}
%
%
%
\title{Relativistic MOND as an alternative to the dark matter paradigm}
%
%
\author{Jacob D. Bekenstein}\ead{bekenste@vms.huji.ac.il}\ead[url]{http://www.phys.huji.ac.il/~bekenste/}
\address{Racah Institute of Physics,
The Hebrew University of Jerusalem\\
Givat Ram, Jerusalem 91904 ISRAEL}
\begin{abstract}
Milgrom's Modified Newtonian dynamics (MOND) provides an efficient way to summarize phenomenology of galaxies which does not lean on the notion of dark matter; it has great predictive power. Here I briefly review MOND as well as its implementation as a nonrelativistic modified gravity theory, AQUAL.  Gravitational lensing and cosmology call for a relativistic gravity theory different from general relativity if dark matter is to be avoided.  In recent years such a theory, T$e$V$e$S, has emerged from the marriage of AQUAL with the timelike vector field of Sanders.  I discuss its structure and some of its successes and shortcomings. 
\end{abstract}
\end{frontmatter}
%
\section{Introduction}
\label{intro}

As other talks in this conference attest, the dark matter (DM) paradigm is a dominant one in physical science, on Earth and away from it.  So why look for alternatives to it?  There are three good reasons.  First, although many astronomical observations seem to demand the existence of DM in large quantities, this is always inferred through the standard law of gravity.  Thus if our understanding of gravity on astronomical scales is flawed, that inference is at grave risk.  Second, apart from sporadic reports which often clash with known constraints, DM has not been identified directly, i.e., by means of a nongravitational experiment.  Third, history suggests that it is best to arrive at the accepted picture of a phenomenon by confronting rather different paradigms for it; confining all attention to the DM paradigm is dangerous until we understand more about the problem it is supposed to solve.

In this review of the MOND alternative to DM, I proceed from some empirical facts about galaxies to the MOND paradigm, and thence to modified gravity theories.

Disk galaxies rotate;   the stars and gas move together in circular concentric orbits.  The circular velocity is best deduced from the Doppler shift of radio waves coming from atomic hydrogen gas, or that of optical lines from the stars.   When plotted against radius $r$, this  velocity constitutes the rotation curve of the galaxy.   According to a simple, Newtonian, balance of centrifugal acceleration and gravitation pull of the galaxy, the rotation curve should first rise (because more mass is being included with increasing radius) and then decrease as $1/\surd r$ .  But the rotation curves of real galaxies, after first rising, almost invariably flatten out, and this behavior continues well beyond the edge of the optical disk as traced by the 21 cm line of atomic hydrogen. 
The reticence of rotation curves to fall off as theoretically expected counts as one of the principal reasons for the widespread belief that disk galaxies are all embedded  in massive DM halos.

And what sets the velocity in the flat part of the rotation curve? A simple rule, the famous and extremely useful Tully-Fisher law.    In one modern version of it,  the infrared (K-band or 2.2 micrometer) luminosity of a disk galaxy is proportional to the fourth power of the circular velocity in the flat part~\cite{Sanders:2002pf}.  In its ``baryonic Tully-Fisher'', version  the mass in baryons (stars and in gas) is proportional to the fourth power of the circular velocity; specifically $M/M_\odot\approx 50 (v/\textrm{km s}^{-1})^4$~\cite{McGaugh:2005qe}.  In brief, the velocity is tied directly to the \textit{visible} matter content.

In the DM halo paradigm such a law must follow from considerations of galaxy formation, a very messy process by all accounts.  It has been hard to understand the sharpness of the correlation in this view.  And the specific form of the power-four law is not well understood either. 

In the large clusters of galaxies,  agglomerations of hundreds of galaxies moving with random velocities of up to $10^3\, \textrm{km\ s}^{-1}$ through a medium of very hot gas, Newtonian virial theorem estimates of a cluster's mass from the galaxy velocities are very large compared to the mass seen directly as galaxies and gas.   This mass discrepancy also shows up when the overall cluster mass is determined Newtonially from the assumption that the hot gas is in a hydrostatic state, or when gravitational lensing by a cluster is analyzed in the framework of general relativity (GR).  The conventional solution is to assume that the typical cluster contains DM to the tune of about five times the visible mass~\cite{Sanders:2002pf}.

We shall set $c=1$ in most equations.

\section{The MOND paradigm and AQUAL}
\label{sec:MOND}

Milgrom's MOND paradigm~\cite{Milgrom:1983ca,Milgrom:1983pn} grew out of dissatisfaction with the DM idea.  Milgrom replaces the standard relation between acceleration of a test particle and the ambient Newtonian gravitational field, $m\vec{a} =-\vec{\nabla}\Phi_{\rm N}$, by 
\begin{equation}
\tilde\mu(|\vec{a}|/a_0)\,\vec{a}=-\vec\nabla\Phi_{\rm N}.
\label{MOND}
\end{equation}
where $\Phi_{\rm N}$ is ascribed to the visible matter alone.
Here $a_0\approx 1.2\cdot 10^{-10}\,  \textrm{m}\, \textrm{s}^{-2}$, a preferred acceleration, is  of the order of the centripetal accelerations of stars and gas clouds in the outskirts of disk galaxies.  The $\tilde\mu(x)$ is a positive function with  $\tilde \mu\to 1$ when $\vec{a}\gg a_0$,  so that we  get back Newtonian behavior as appropriate in the solar system where $|\vec{a}|\sim 10^{-2}\, \textrm m\, s^{-2}$, and $\tilde \mu(x)\approx x$ when $\vec{a}\ll a_0 $.  This last is the deep MOND limit which should be relevant in the outer parts of disk galaxies.  If for it we set $|\mathbf {a}|=v^2/r$ and $|\vec{\nabla}\Phi_{\rm N}|= GM/r^2$ ($M$ total galaxy mass), we find the MOND formula to be satisfied for $v=$const.  Thus in the outskirts of a disc galaxy, where mass no longer grows with radius, MOND predicts a flat rotation curve.  And from the coefficients in the above we get that $M=v^4/G a_0$, which is just the baryonic Tully-Fisher law.  Because the mass to infrared luminosity is fairly constant among disk galaxies~\cite{Sanders:2002pf}, the infrared luminosity version of the Tully-Fisher law follows too.  One can view $a_0$ as determined by the observed Tully-Fisher law.

How well does the MOND paradigm reproduce the shapes of observed  disc galaxy rotation curves in terms of the mass distribution actually seen in them (gas and stars)?   Fits to over one hundred disk galaxies bear witness that MOND is very successful~\cite{Sanders:2002pf,Sanders:2007rg,Begeman:1991iy}.  With the choice $\tilde\mu(x)=x(1+x)^{-1}$ (the simple function~\cite{Binney:2005fg}) and with the mass-to-luminosity ratio $\Upsilon$ for the stellar population as its only fitting parameter, MOND is as successful as DM halo models with three fitting parameters (typically $\Upsilon$, central velocity dispersion of DM particles and scale of length of halo).  In addition, the features in a MOND predicted rotation curve correlate well with features in the visible mass distribution of the relevant galaxy.   This is embarrasing for the DM paradigm in which the gravitational field that  controls the rotation curve comes mostly from the DM.  It is no longer controversial that for disk galaxies of all scales, MOND provides the most economical description.

While the MOND equation (\ref{MOND}) is incompatible with the conservation laws (except if applied solely to test particles), the MOND paradigm can be made  dynamically consistent by reformulating it as a modified gravity theory, AQUAL, derived from an \textit{AQU}adratic \textit{L}agrangian~\cite{Bekenstein:1984tv}.
In AQUAL the Poisson equation is replaced by 
\begin{equation}
\vec\nabla\cdot[\tilde\mu(|\vec\nabla\Phi|/a_0)\vec\nabla\Phi]=4\pi G\rho,
\label{AQUAL}
\end{equation}
where $\rho$ is the baryonic matter mass density, and $-\vec\nabla\Phi$ gives the acceleration field.
For spherically symmetric systems the first integral of the AQUAL equation with $\vec\nabla\Phi\to -\vec{a}$ reproduces Eq.~(\ref{MOND}) exactly.  For disk like systems the two differ by some 10-15\%~\cite{Milgrom:1986ib}.  Thus given the notoriously inaccurate character of astronomical data, AQUAL fits disk galaxies just as well as does the MOND equation. 

Not only is AQUAL equipped with the usual set of conservation laws, it also removes the MONDian paradox of why stars whose entrails are mostly strongly accelerated ions move in a galaxy in non-Newtonian fashion~\cite{Bekenstein:1984tv}.  AQUAL has established~\cite{Bekenstein:1984tv} the long conjectured external field effect~\cite{Milgrom:1983ca}: the suppression of MOND effects in a system immersed in a sufficiently strong gravity field.  And it has permitted full understanding of how MOND stabilizes galaxies against the ``cold disk'' instability~\cite{Milgrom:1989pr,Brada:1999cc} which supplied the original  $raison\ d'\hat etre$ for DM halos.  Today AQUAL is the basis for several $N$-body simulators
which are elucidating the role of MONDian effects in galactic formation and evolution~\cite{Tiret:2007cd,Nipoti:2007ab}. (For more on MOND and AQUAL see recent reviews, e.g.~\cite{Bekenstein:2006ab,Milgrom:2008rv}).

But AQUAL is powerless to give a complete account of the dynamics of the large clusters of galaxies. In one of these the accelerations of the galaxies near the center can well exceed $a_0$, so both MOND formula and AQUAL would predict nearly Newtonian behavior for cluster cores.  Thus while MOND-AQUAL models of cluster dynamics explain some of the mass discrepancy, they leave a factor of 2-3 unexplained~\cite{Sanders:2002pf}.  Unlike galaxies, clusters seem to require invisible  mass.  The promising idea~\cite{Sanders:2003ap} that the invisible mass is made up of neutrinos (which \textit{are} known to exist)  has been running into problems recently~\cite{Zhao:2008na}. Perhaps clusters contain large amounts of dark baryons~\cite{Milgrom:2008tz}. Or the problem might be solved in MOND spirit if a natural way were found to shift upwards Milgrom acceleration scale $a_0$ in systems bigger than galaxies.

Nonrelativistic by construction, AQUAL cannot serve to describe gravitational lensing by galaxies or clusters of galaxies, or cosmology, both essentially relativistic in nature, and strongly bound up with the DM issue.  In fact, just a few years ago it was common for DM pundits to reject  MOND  because ``it cannot be framed relativistically''.  There were indeed various stumbling blocks on the way to this goal, but they proved surmountable.  Today there are an handful of relativistic MOND gravity theories which disprove that pessimistic assessment~\cite{Bekenstein:2004ab,Sanders:2005vd,Zlosnik:2007lq}.  We turn to the earliest and perhaps simplest of these.

\section{Tensor-Vector Scalar theory and gravitational lensing}

The requisite relativistic theory of gravity must depart from GR---the standard gravity theory today---because the latter has a uniformly Newtonian nonrelativistic limit.  But it should not depart from it too strongly because GR is known to be accurate at solar system scale.  GR embodies the equivalence principle; in its weak form this principle requires that all matter propagate on the same metric.  Because of its philosophical basis and strong support from Earth-bound experiments, we would like to retain that feature.  This means all matter actions in the new theory must be written with one metric, $\tilde g_{\alpha\beta}$.   If we write the gravitational action in Hilbert-Einstein form using $\tilde g_{\alpha\beta}$ to form the scalar of curvature, $R$, we are back to GR (unless we are willing to introduce new types of gravitational fields).  We may choose to depart from GR by modifying the gravitational action, e.g., as in $f(R)$ theories.  But whatever successes maybe claimed by these last in regard to the dark energy puzzle, they cannot reproduce the regularities shown by galaxies and summarized by MOND.  

By contrast, T$e$V$e$S retains the Einstein-Hilbert gravitational action, but written with a metric $g_{\alpha\beta}$ distinct from
$\tilde g_{\alpha\beta}$; this gives rise a controlled departure from GR.  Relating  $g_{\alpha\beta}$ and  $\tilde g_{\alpha\beta}$ by a conformal factor is a simple possibility.  This leads to a relativistic version of AQUAL~\cite{Bekenstein:1984tv}.  However, the scalar field is found to be afflicted by superluminal propagation, and the theory does not provided for the enhanced gravitational lensing that would be required to dispense altogether with DM in galaxies.

A significant idea here is Sanders' unit timelike 4-vector field $\mathcal{ U}_\alpha$~\cite{Sanders:1997tq}.  Instead of the conformal relation between the metrics, which is the root of relativistic AQUAL's failure in the gravitational lensing domain, Sanders used
\begin{equation}
 \tilde g_{\alpha\beta}=e^{-2\phi}\,  g_{\alpha\beta}- (e^{2\phi}-e^{-2\phi})\, \mathcal{ U}_\alpha\,\mathcal{ U}_\beta
 \label{gtilde}
 \end{equation}
($\phi$ is a dimensionless  auxiliary scalar field), which \textit{can} give lensing of the correct strength.  But since Sanders regarded $\mathcal{ U}_\alpha$ as constant and pointed in the time direction of the cosmos, the resulting theory was not covariant.  

T$e$V$e$S~\cite{Bekenstein:2004ab} achieves covariance by converting $\mathcal{ U}_\alpha$ into a dynamical field with action
 \begin{equation}
S_v = -\frac{ 1}{32\pi G}\int
\Big[K g^{\alpha\beta}g^{\mu\nu}  \mathcal{ U}_{[\alpha,\mu]} \mathcal{ U}_{[\beta,\nu]}  
-2 \lambda(g^{\mu\nu}\mathcal{ U}_\mu\mathcal{ U}_\nu +1)\Big](-g)^{1/2} d^4 x,
\end{equation}
where $K$ ia dimensionless parameters and $\lambda$ is a Lagrange multiplier field charged with maintaining the normalization of  $\mathcal{ U}_\alpha$ to $-1$, which is the easiest way to make the vector everywhere timelike.  (Of late it has been found useful to buttress $S_v$ with a term $\bar K ( g^{\alpha\beta}  \mathcal{ U}_{\alpha;\beta})^2$ to prevent the formation of caustics of the integral lines of $g^{\alpha\beta}\mathcal{ U}_\beta$ and inconsistencies in the weak field approximation to the theory).

The action governing this scalar's evolution is
\begin{equation}
\label{bekscalaraction}S_s=-\frac{1}{2 k^2 \ell^2 G}\int
\mathcal{F}\big(k
\ell^2 ( g^{\alpha\beta}-g^{\alpha\gamma}g^{\beta\delta}\,\mathcal{ U}_\gamma\, \mathcal{ U}_\delta)\phi_{,\,\alpha}\phi_{,\,\beta}\big)\,(-g)^{1/2} d^4 x,
\end{equation}
where $\mathcal{F}$ is a positive function, $k$ is another dimensionless coupling constant, and $\ell$ is a constant scale of length.  This differs from a covariant version of the AQUAL lagrangian~\cite{Bekenstein:1984tv} only in the appearance of $g^{\alpha\beta}-\mathcal{ U}^\alpha\mathcal{ U}^\beta$ instead of $g^{\alpha\beta}$, a change introduced to forestall superluminal $\phi$ propagation.   In the limit $k\to 0$ with $K\propto k$ and $\ell\propto k^{-3/2}$,  T$e$V$e$S reduces to GR~\cite{Bekenstein:2004ab}.  More on T$e$V$e$S can be found in the reviews~\cite{Bekenstein:2006ab,Bekenstein:2009kh}.

For weak $\Phi_{\rm N}$ and $\phi$  and a quasistatic situation, T$e$V$e$S yields the linearized line element (corresponding to $\tilde g_{\alpha\beta}$)
\begin{equation}
d\tilde s^2=-(1+2\Phi_{\rm N}+2\phi)dt^2+(1-2\Phi_{\rm N}-2\phi)(dx^2+dy^2+dz^2).
\label{met2}
\end{equation} 
The scalar $\phi$'s equation takes on the form of Eq.~\ref{AQUAL} sourced by the same $\rho$ that sources $\Phi_{\rm N}$, with  $\mu(y)\equiv d\mathcal{F}(y)/dy$   playing the role of $\tilde\mu$.  With the choice  $\mu(y)=D \surd y$ for $0<y\ll1$ ($D$ a positive constant),  and $\mu(y)\to 1$ for $y\to\infty$,  the role of $a_0$  is played by   $c^2\surd k(4\pi D \ell)^{-1}$   Whenever $|\vec\nabla\phi |\ll a_0$, the scalar dominates over $\Phi_{\rm N}$ in the metric, and we have AQUAL behavior of dynamics with $\Phi\approx \phi$.  
And whenever  $|\vec\nabla\phi |\gg a_0$, $\phi$ has the same form as $\Phi_{\rm N}$, so that T$e$V$e$S reduces to Newtonian gravity, albeit with a rescaled gravitational constant.     In these senses  T$e$V$e$S is a relativistic MOND gravity.    

According to Eq.~\ref{met2}, in linearized approximation in T$e$V$e$S, just as in GR, one and the same potential rules dynamics and controls gravitational lensing.  If T$e$V$e$S describes well the dynamics of a system \textit{everywhere},  it should predict the same pattern of gravitational lensing that GR would predict, were the latter supplemented with the appropriate distribution of DM to fit the dynamics everywhere.  In practice this obvious dictum may work less than perfectly when the extant data pertains to just a limited part of the system.  T$e$V$e$S does well in the confrontation with data for strong lensing (multiple images) by galaxies~\cite{Chen:2008ab,Bekenstein:2009kh,Bekenstein:2006ab}.  It meets problems for weak lensing (distorted single images) by clusters of galaxies~\cite{Takahashi:2007ty,Feix:2008ab,Bekenstein:2009kh}; basically it does not account for the observed distortion without the help of invisible matter apart (in addition a reasonable dose of neutrinos).  This problem is dramatized by a handful of colliding clusters, and has been made much of.  But lamenting over T$e$V$e$S would be premature.  MOND has never dealt perfectly with the dynamics of clusters.  So T$e$V$e$S, which was designed with MOND in mind, could not expected to do well in this business, and modifications of it may be in order.  And, as already mentioned, clusters may well contain large amounts of \textit{as yet} invisible matter~\cite{Milgrom:2008tz}.  Enlightenment  from the MOND paradigm is not at an end!

%

%
\end{document}